\documentclass[sigplan,nonacm]{acmart}

\graphicspath{ {./fig/} }
\usepackage{units} 
\usepackage{subcaption} 
\usepackage{varioref} 
\usepackage{listings} 
\usepackage{cancel} 

\newcommand{\real}{\mathbb{R}} 
\newcommand{\positive}{\mathbb{N}} 
\newcommand{\degree}{\text{\textdegree}} 
\newcommand{\tran}{^\mathsf{T}} 
\DeclareMathOperator*{\clamp}{clamp} 
\DeclareMathOperator{\asin}{arcsin} 
\DeclareMathOperator{\atan}{arctan} 
\allowdisplaybreaks[1]

\newcommand{\noun}[1]{\textsc{#1}} 

\theoremstyle{remark}
	\newtheorem*{note}{Note}
	\newtheorem*{rem}{Remark}

	\newtheorem*{problem*}{Problem}
	\newtheorem*{prop*}{Proposition}
	
	\newtheorem*{example*}{Example}
\theoremstyle{definition}

\theoremstyle{plain}

\lstdefinelanguage{GLSL}
{
	morekeywords={
		attribute,const,uniform,varying,
		break,continue,do,
		for,while,
		switch,case,default,
		if,else,
		in,out,inout,
		float,int,void,bool,true,false,
		discard,return,
		mat2,mat3,mat4,
		mat2x2,mat2x3,mat2x4,
		mat3x2,mat3x3,mat3x4,
		mat4x2,mat4x3,mat4x4,
		vec2,vec3,vec4,
		ivec2,ivec3,ivec4,
		bvec2,bvec3,bvec4,
		uint,uvec2,uvec3,uvec4,
		lowp,mediump,highp,precision,invariant,
		sampler1D,sampler2D,sampler3D,samplerCube,
		struct,
		define,undef,
		ifdef,ifndef,elif,endif,
	},
	sensitive=false, 
	morecomment=[l]{//},
	morecomment=[s]{/*}{*/},
	morestring=[b]"
}
\received{February 2021}
\received{August 2021}
\received{December 2023}
\received{February 2024}

\begin{document}
\title{Aximorphic Perspective Projection Model for Immersive Imagery}

\author{Jakub Maksymilian Fober}
\orcid{0000-0003-0414-4223}
\email{jakub.m.fober@protonmail.com} 
\email{talk@maxfober.space}

\renewcommand\shortauthors{Fober, J.M.}

\begin{abstract}
	A wide choice of cinematic lenses enables motion-picture creators to adapt image visual-appearance to their creative vision.
	Such choice does not exist in the realm of real-time computer graphics, where only one type of perspective projection is widely used, a linear perspective.
	This paper presents an extended perspective imaging model, which can represent distortion and FoV parameters of entire variety of film and photographic lenses (e.g., wide-angle, fisheye, anamorphic), while preserving parametrization in an artistically convincing manner.
	Self-experimentation with the model revealed that each projection type provides accurate perception of a different aspect of depicted space (e.g., speed, distance, shape).
	Presented model, enables combination of multiple projections, each on a different axis of the image, to achieve optimal perception for a given scenario.
	This new projection, named \emph{aximorphic}, was made available here, under an open license (CC BY-SA 3.0), for a wide and easy adoption.
\end{abstract}

\ccsdesc[500]{Computing methodologies~Perception}
\ccsdesc[500]{Computing methodologies~Ray tracing}
\ccsdesc[300]{Human-centered computing}
\ccsdesc[300]{Applied computing~Media arts}


\keywords{Curvilinear Perspective, Panini, Anamorphic, Cylindrical, Fisheye, lens Map}

\maketitle


\begin{figure}[bh]
	\footnotesize
	\copyright\ 2021-\the\year\ Jakub Maksymilian Fober\smallskip\\
	\href{https://creativecommons.org/licenses/by-sa/3.0/}{\includegraphics{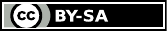}}\\
	This work is licensed under Creative Commons BY-SA 3.0 license.
	\url{https://creativecommons.org/licenses/by-sa/3.0/legalcode}\\
	For all other options including custom licensing, contact the owner/author(s).
\end{figure}


\section{Introduction}

The perspective in the computer real-time graphics hasn't changed since the dawn of CGI. It is based on a concept as old as the fifteenth-century Renaissance, a linear projection \cite{Alberti1970OnPaining,Argan1946BrunelleschiPerspective,McArdle2013VisualSphereEuclid}. This situation is similar to the beginnings of photography, where only one type of lens was widely used, an Anno Domini 1866 \emph{Rapid Rectilinear} \cite{Kingslake1989RapidRectilinearLens} lens. Linear perspective even at the time of its advent, 500 years ago, has been criticized for distorting proportions \cite{DaVinci1632TeatriseOnPainting}. In a phenomenon known today as \emph{Leonardo's paradox} \cite{Dixon1987Paradox}, in which figures further away but at the periphery appear larger, than those located near the optical center. Computer graphics really skipped the artistic achievements of the last five centuries in regard to perspective. This includes the cylindrical perspective of Pannini \cite{Sharpless2010Panini}, Barker \cite{Wikipedia2019BarkerPanorama}, and anamorphic lenses used in cinematography \cite{Sasaki2017_5PilarsOfAnamorphic,Sasaki2017Anamorphic,Giardina2016AnamorphicUltra70,Neil2004CineLens}. The situation is even more critical, as there is no mathematical model for generating anamorphic projection in an artistically-convincing manner \cite{Yuan2009AnamorphicLens}.
Some attempts were made at alternative projections for computer graphics, with fixed cylindrical or spherical geometry \cite{Sharpless2010Panini,Baldwin2014PerspectiveComparisonTests}. A parametrized perspective model was also proposed as a new standard \cite{Correia2007ExtendedPerspectiveSystem}, but wasn't adopted. It included interpolation states in between rectilinear/equidistant and spherical/cylindrical projection. The cylindrical parametrization of this solution was merely an interpolation factor, where intermediate values did not correspond to any common projection type. Therefore, it was not well-suited for artistic or professional use.

\subsection[Axiom of Pyramid Frustum]{Break from Axiom of Pyramid Frustum}
The notion of sphere as a projective surface that incorporates cartographic mapping to produce a perspective picture became popularized \cite{Penaranda2015SphereInterpolation,German2007PanoramaOnFlatSurface,Williams2015QuakeBlinky}. Also, perspective parametrization that transitions according to the content (by the view-tilt angle) has been developed, as a modification to the computer game \noun{Minecraft} \cite{Williams2017Minecraft}. But the results of these solutions were more a gimmick and have not found practical use. The linear perspective projection was still the way-to-go for most digital content. Some state-of-the-art video games incorporated limited lens distortion, like \noun{Resident Evil} series (after 2017), \emph{Alan Wake 2} and more strongly \emph{Unrecord}.

One of the reasons for a limited adoption of a non-linear projection was the fixed-function architecture of GPUs in regard to rasterization. But with the advent of real-time ray-tracing and variable shading-rate, more exotic projections could become widely adopted and integrated into the tools.

\subsection{Presented New Model}
This paper aims to provide a perspective model with a mathematical parametrization that allows artistic-style interaction with image geometry. Similar in a way film directors choose lenses for each scene based on aesthetics \cite{Giardina2016AnamorphicUltra70,Sasaki2017Anamorphic,Neil2004CineLens}, but with parametrization that offers a greater degree of control.
It also provides a psycho-physiological correlation between perspective model parameters and the perception of depicted space attributes, like distance, size, proportions, shape or speed. That mapping enables the use of the presented model in a professional environment, where image geometry is specifically suited for a task \cite{Whittaker1984StereographicCrystal}.

The presented perspective model is named \emph{aximorphic} (from \emph{axis-}, line of symmetry, and \emph{+morph\=e}, shape; ``varying shape across axes'').


\subsection[Naming Convention]{Document Naming Convention}

This document uses the following naming convention:
\begin{itemize}
	\item A left-handed coordinate system is used.
	\item Vectors are presented in column format.
	\item Matrices use row-major order and are denoted as  ``$M_{\text{row}\,\text{col}}$''.
	\item Matrix multiplication is denoted as ``$[\text{column}]_a\cdot[\text{row}]_b=M_{a\,b}$''.
	\item A single bar enclosure ``$|u|$'' represents the absolute value of a scalar.
	\item A single bar enclosure ``$|\vec v|$'' represents the length of a vector.
	\item Vectors with an arithmetic sign, or without, are calculated component-wise to form another vector.
	\item Centered dot ``$\cdot$'' represents the dot product of two vectors.
	\item Square brackets with a comma ``$[f,c]$'' denotes an interval.
	\item Square brackets with blanks ``$[x\ y]$'' denotes a vector or a matrix.
	\item The power of ``$^{-1}$'' implies the reciprocal of the value.
	\item The \emph{QED} symbol ``$\square$'' marks the final result or output.
\end{itemize}
This naming convention simplifies the process of translating formulas into shader code.

\section[Primary-ray map]{Aximorphic Primary-ray Map}
\label{sec:primary-ray map}

\begin{figure*}
	\begin{subfigure}[t]{2in}
		\includegraphics{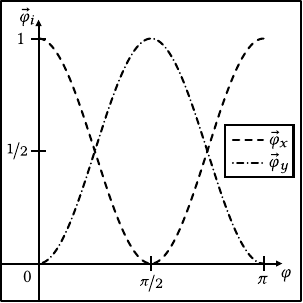}
		\caption{Graph illustrating mapping of the angle $\varphi$, to aximorphic interpolation weights $\vec\varphi_x$ and $\vec\varphi_y$. Illustrates circular distribution of $\vec\varphi$, in a periodic function.}
		\label{fig:circular distribution}
	\end{subfigure}
	\hfill
	\begin{subfigure}[t]{2in}
		\includegraphics{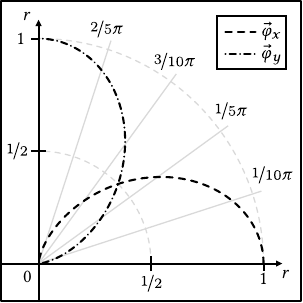}
		\caption{Radial graph illustrating mapping of aximorphic interpolation weights $\vec\varphi_x$ and $\vec\varphi_y$ (as radius $r$), to angle $\varphi$ in radians. Such that $\vec\varphi_x+\vec\varphi_y=1$.}
		\label{fig:philerp to r}
	\end{subfigure}
	\hfill
	\begin{subfigure}[t]{2in}
		\includegraphics{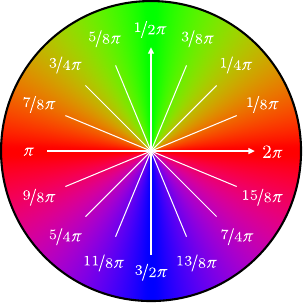}
		\caption{Radial graph illustrating aximorphic interpolation vector $\vec\varphi\in[0,1]^2$. Here colors \emph{red}, \emph{green}, \emph{blue} represent contribution of $\vec k_x$, $\vec k_y$ and $\vec k_z$ respectively.}
		\label{fig:philerp distribution}
	\end{subfigure}
	\caption{Illustrating correlation between the aximorphic interpolation weights $\vec\varphi_x$, $\vec\varphi_y$ and the spherical angle $\varphi$.}
	\label{fig:phi correlation}
\end{figure*}

If we assume that a projective visual space is spherical \cite{McArdle2013VisualSphereEuclid,Fleck1994FisheyeEquations}, one can define perspective picture as an array of rays pointing to the surface of visual sphere. This is how the algorithm described below will output a viewing-ray map (aka lens map). Lens map is a two-dimensional $\real^3$ vector field representing viewing rays, where each ray is assigned to a screen pixel. Visual sphere as the image model enables wider angle of view, beyond the limit for planar projection of 180\textdegree. Such vector field can be easily converted to a cube UV map, ST map, or other screen distortion format.

Here, the procedural algorithm for lens map uses two types of input values from the user; distortion parameter for two principal axes and \emph{focal-length} or \emph{angle-of-view} (aka FoV). Two distinct principal axes define the aximorphic projection type. Each axis distortion profile is expressed by the azimuthal projection factor $k$ \cite{Krause2019PTGuiFisheye_k-factor,Bettonvil2005FisheyeEquations,Fleck1994FisheyeEquations}. Both principal axes share the same focal-length value $f$.
The evaluation of each principal-axis distortion profile produces spherical angles $\vec\theta_x$ and $\vec\theta_y$. These angles are then combined to form the aximorphic azimuthal-projection angle $\theta'$. The interpolation of $\vec\theta$ components is achieved through aximorphic weights $\vec\varphi_x$ and $\vec\varphi_y$, which are derived from the spherical angle $\varphi$ of the azimuthal projection.
\begin{note}
Calculation of $\varphi$ angle is omitted here, as view-coordinate $\vec v$ alone allows for direct calculation of $\vec\varphi$ weights.
\end{note}

\begin{subequations}
\label{eq:double spherical theta}
\begin{align}
	r &= |\vec v| = \sqrt{\vec v^2_x+\vec v^2_y}
\\
	\vec\theta_{x,y} &=
		\begin{cases}
			\frac{\atan\big( \frac{r}{f}\vec k_{x,y} \big)}{k_{x,y}}, & \text{if }\vec k_{x,y}>0 \\
			\frac{r}{f}, & \text{if }\vec k_{x,y}=0 \\
			\frac{\asin\big( \frac{r}{f}\vec k_{x,y} \big)}{k_{x,y}}, & \text{if }\vec k_{x,y}<0 \\
		\end{cases}
\\
	\begin{bmatrix}
		\vec\varphi_x \\
		\vec\varphi_y
	\end{bmatrix}
	&=
	\begin{bmatrix}
		\cos^2\varphi \\
		\sin^2\varphi
	\end{bmatrix}
	=
	\begin{bmatrix}
		\frac{1}{2}+\frac{\cos(2\varphi)}{2} \smallskip\\
		\frac{1}{2}-\frac{\cos(2\varphi)}{2}
	\end{bmatrix}
	=
	\begin{bmatrix}
		\frac{\vec v^2_x}{\vec v^2_x+\vec v^2_y} \smallskip\\
		\frac{\vec v^2_y}{\vec v^2_x+\vec v^2_y}
	\end{bmatrix}
	,
\end{align}
\end{subequations}
where $r\in\real_{>0}$ is the view-coordinate radius (vector magnitude). Vector $\vec\theta\in[0,\pi]^2$ contains two incidence angles (measured from the optical axis) of two azimuthal projections determined by two distinct $k$ parameters. Vector $\vec\varphi\in[0,1]^2$ contains the aximorphic interpolation weights, which are linear $\vec\varphi_x+\vec\varphi_y=1$, but exhibit spherical distribution (see Figure \vref{fig:phi correlation}). Vector $\vec k\in[-1,1]^2$ (also $[-1,1]^3$ in a variant from subsection \ref{sec:asymmetry}) describes two power axes of aximorphic projection.
The algorithm is evaluated per-pixel for position $\vec v\in\real^2$ in view-space coordinates, centered at the optical axis and normalized at the chosen angle-of-view (horizontal or vertical).
The final aximorphic incidence angle $\theta'$ is obtained through interpolation of $\vec\theta$ components by $\vec\varphi$ weights.
\begin{subequations}
\label{eq:incidence vector}
\begin{align}
	\theta'
	&=
	\begin{bmatrix}
		\vec\theta_x \\
		\vec\theta_y
	\end{bmatrix}
	\cdot
	\begin{bmatrix}
		\vec\varphi_x \\
		\vec\varphi_y
	\end{bmatrix}
	=
	\vec\theta_x\vec\varphi_x
	+
	\vec\theta_y\vec\varphi_y
\\
	\begin{bmatrix}
		\hat G_x \\
		\hat G_y \\
		\hat G_z
	\end{bmatrix}
	&=
	\begin{bmatrix}
		\sin\theta'
		\begin{bmatrix}
			\cos\varphi \\
			\sin\varphi
		\end{bmatrix} \\
		\cos\theta'
	\end{bmatrix}
	=
	\begin{bmatrix}
		\frac{\sin\theta'}{r}
		\begin{bmatrix}
			\vec v_x \\
			\vec v_y
		\end{bmatrix} \\
		\cos\theta'
	\end{bmatrix}, \qed
\end{align}
\end{subequations}
here $\theta'\in(0,\pi]$ is the aximorphic incidence angle, measured from the optical axis. This measurement resembles azimuthal projection of a globe (here a visual sphere) \cite{McArdle2013VisualSphereEuclid}.
The final incidence vector $\hat G\in[-1,1]^3$ (aka viewing-ray) is obtained from the aximorphic angle $\theta'$.
Parameters $r,\vec v,\vec\varphi$ are in view-space, while $\vec\theta,\theta',\varphi,\hat G$ are in visual-sphere space.
Essentially, the aximorphic primary-ray map preserves the azimuthal angle $\varphi$ while modulating only the radius of the image.

\paragraph{Inverse Mapping} It can be obtained through a lookup mesh shaped as a tessellated screen-plane, mapped to the visual-sphere surface by equation \eqref{eq:incidence vector}. This mesh can be directly sampled by a ray to read the UV coordinates, or rasterized to a UV coordinates map. Ray-mesh sampling offers the advantage of an unlimited angle-of-view compared to texture sampling, which is practically limited to around $140\degree\Omega$.

\subsection[Focal Length and AOV]{Focal Length and Angle-of-View}
To enhance control over the image, a mapping between the angle-of-view $\Omega$ and the focal length $f$ can be established. Here, the focal length is expressed in reciprocal format to optimize its use in equation \eqref{eq:double spherical theta}.
\begin{equation}
\label{eq:focal length}
	f^{-1} =
		\begin{cases}
			\frac{\tan\big( \frac{\Omega_h}{2}\vec k_x \big)}{r\vec k_x}, & \text{if }\vec k_x>0 \\
			\frac{\Omega_h}{2r}, & \text{if }\vec k_x=0 \\
			\frac{\sin\big( \frac{\Omega_h}{2}\vec k_x \big)}{r\vec k_x}, & \text{if }\vec k_x<0,
		\end{cases}
\end{equation}
where $\Omega_h\in(0,\tau]$ denotes horizontal angle of view, and $r$ denotes radius at $\Omega$. Similarly, vertical $\Omega_v$ can be obtained using $\vec k_y$ parameter instead. The resultant value $\nicefrac{1}{f}\in\real_{>0}$ is the reciprocal focal-length.

\begin{rem}
	The focal-length $f$ value must be the same for both $\vec\theta_x$ and $\vec\theta_y$. Therefore, only one reference angle $\Omega$ can be chosen, either horizontal or vertical.
\end{rem}

Inverse function, to equation \eqref{eq:focal length}, for angle-of-view $\Omega$, from focal-length $f$, is obtained as follows:
\begin{equation}
\label{eq:vertical AOV}
	\Omega_v =
		\begin{cases}
			2\frac{\atan \big(\frac{r}{f}\vec k_y \big)}{\vec k_y}, & \text{if }\vec k_y>0 \\
			2\frac{r}{f}, & \text{if }\vec k_y=0 \\
			2\frac{\asin \big(\frac{r}{f}\vec k_y \big)}{\vec k_y}, & \text{if }\vec k_y<0,
		\end{cases}
\end{equation}
This formula can be used to obtain the actual vertical angle-of-view from a horizontally established focal length. Similarly horizontal angle $\Omega_h$ can be obtained using the $\vec k_x$ parameter. Here input value $r$ denotes radius at $\Omega$.
\begin{table*}
	\begin{subtable}[t]{\columnwidth}
		\centering
		\begin{tabular}{rcl}
			\toprule
			\multicolumn{2}{l}{Value of $k$} & Azimuthal projection type \\ \midrule
			$k_i=$ & $1$                & Rectilinear (Gnomonic) \\
			$k_i=$ & $\nicefrac{1}{2}$  & Stereographic \\
			$k_i=$ & $0$                & Equidistant \\
			$k_i=$ & $-\nicefrac{1}{2}$ & Equisolid \\
			$k_i=$ & $-1$               & Orthographic (azimuthal) \\
			\bottomrule
		\end{tabular}\\
		\smallskip
		\footnotesize\emph{Source:} PTGui 11 fisheye factor \cite{Krause2019PTGuiFisheye_k-factor}.
		\bigskip
		\caption{Primary $k$ values and corresponding azimuthal projection type.}
		\label{tab:k values}
	\end{subtable}
	\hfill
	\begin{subtable}[t]{\columnwidth}
		\centering
		\begin{tabular}{ll}
			\toprule
			Azimuthal projection type & Perception of space \\\midrule
			Rectilinear               & straight lines \\
			Stereographic             & shapes, angles \\
			Equidistant               & speed, aim position \\
			Equisolid                 & distances, sizes \\
			Orthographic (azimuthal)  & brightness (no vignetting) \\
			\bottomrule
		\end{tabular}\\
		\smallskip
		\footnotesize\emph{Source:} Properties of azimuthal projections and empirical self study using various video games with shader overlay, by ReShade shader injection.
		\bigskip
		\caption{Azimuthal projection type and corresponding enhancement for a given space attribute perception.}
		\label{tab:perception}
	\end{subtable}
	\caption{Tables presenting perspective parameters, corresponding projection type and associated perception attitude.}
	\label{tab:k values and perception}
\end{table*}

\subsection[Asymmetry]{Asymmetrical Aximorphism}
\label{sec:asymmetry}

Parameter $\vec k_y$ can be further augmented to produce an asymmetrical aximorphic projection by introducing a third input value denoted as $\vec k_z$. In such a case, the bottom and top halves of the image can present different azimuthal projections along the axis.
\begin{equation}
\label{eq:asymmetry}
	\vec k_y' =
		\begin{cases}
			\vec k_z, & \text{if }\vec v_y<0 \\
			\vec k_y, & \text{otherwise},
		\end{cases}
\end{equation}
therefore $\vec k_y'$ replaces $\vec k_y$ in equations \eqref{eq:double spherical theta}, \eqref{eq:vertical AOV} and \eqref{eq:vignette mask}.

Asymmetrical aximorphism can be applied to any side of the principal axes. A use case for such a perspective could be in racing, where the bottom half of the screen contains an image of the road. Choosing equidistant projection (which preserves angular speed) would provide an accurate perception of velocity. The top half of the screen contains the image of opponent vehicles or the road ahead. Choosing stereographic projection (which preserves angles and proportions) would provide an enhanced perception for choosing the optimal trajectory. For the horizontal power axis, choosing equisolid projection (which preserves distance) would enhance the perception of distance to the turn for braking.

\subsection{Vignetting Mask}

\begin{figure*}
	\hfill
	\begin{subfigure}[t]{2in}
		\includegraphics{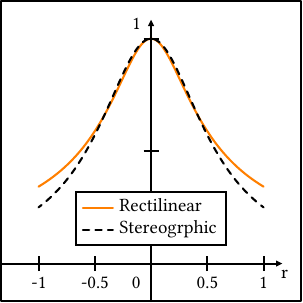}
		\caption{Image radius to vignetting intensity graph, for Rectilinear projection at $140\degree\Omega$ and for Stereographic projection at $240\degree\Omega$.}
		\label{fig:vignetting positive}
	\end{subfigure}
	\hfill
	\begin{subfigure}[t]{2in}
		\includegraphics{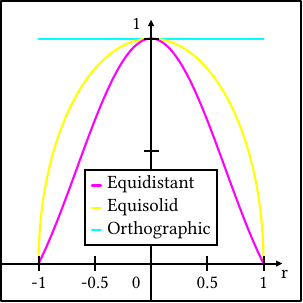}
		\caption{Image radius to vignetting intensity graph, for Equidistant and Equisolid projection at $360\degree\Omega$ and for Orthographic (azimuthal) projection at $180\degree\Omega$.}
		\label{fig:vignetting negative}
	\end{subfigure}
	\hfill
	\hfill
	\caption{Plotting of vignetting intensity across the image radius, for each azimuthal projection at various $\Omega$ angles.}
	\label{fig:vignetting}
\end{figure*}

Vignetting is a crucial visual symbol indicating the stretching of the visual sphere by the projection. Incorporating a vignetting effect with a custom projection enhances spatial perception.

Here, the vignetting mask is obtained in the same way for all aximorphic, anamorphic, and spherical projections. It is generated as the ratio of the circumference of a small circle $(2\pi\sin\theta')$ to the circumference of the image circle $(2\pi r_i)$, where the image circle's radius $r_i$ is obtained through the normalization of the picture-space radius $r$ by the focal length $f$.
\begin{equation}
\label{eq:vignette mask}
	\Lambda =
		\begin{cases}
			1,& \text{if }r=0 \\
			\frac{\sin\theta'}{r}f,& \text{else}, \qed
		\end{cases}
\end{equation}
where $\Lambda\in[0,1]$ is the aximorphic vignetting mask value, which is inversely proportional to the scaling of visual sphere surface in image space.
\begin{note}
	In the shader implementation, conditional branching can be omitted as equation $\frac{\sin0\degree}{0}$ will automatically yield 1.
\end{note}

This vignetting model accounts for the natural falloff due to the stretching of the projected visual sphere. In real optical systems, the vignetting effect is often enhanced at the borders by the gradual occlusion of the aperture, especially at lower f-stops. At lower f-stops, some lens casing elements can block the entrance pupil at steep angles, usually when the aperture is wide-open. Therefore presented model is closest achieved at lowest apertures, when the entrance pupil is small and vignetting is closest to natural.

\begin{table}
	\begin{tabular}{lrcccl}
		\toprule
		Picture content type   & \multicolumn{5}{l}{Aximorphic $\vec k$ values} \\
		\midrule
		Racing                 & $\vec k=[$ & $-\nicefrac{1}{2}$ & $\nicefrac{1}{2}$   & $0$                 & $]$ \\
		Flying                 & $\vec k=[$ & $\nicefrac{1}{2}$  & $0$                 & $-\nicefrac{1}{2}$  & $]$ \\
		First-person (generic) & $\vec k=[$ & $\nicefrac{1}{2}$  & $\nicefrac{22}{25}$ & $\nicefrac{22}{25}$ & $]$ \\
		First-person (aiming)  & $\vec k=[$ & $0$                & $\nicefrac{1}{2}$   & $\nicefrac{1}{2}$   & $]$ \\
		\midrule
		Pan motion                           & \multicolumn{5}{l}{$\vec k_x\neq\vec k_y$} \\
		Roll motion                          & \multicolumn{5}{l}{$\vec k_y=\vec k_x$} \\
		Tilt motion\emph{\textsuperscript a} & \multicolumn{5}{l}{$\vec k_y\rightarrow\vec k_x$} \\
		\bottomrule
	\end{tabular}\\
	\smallskip
	\footnotesize\emph{Source:} Values determined empirically with self experimentation using various competitive video games, in accordance to data in Table \ref{tab:perception}.

	\emph{\textsuperscript a}Mapping of vertical distortion by a tilt motion introduced first in a Minecraft mod \cite{Williams2017Minecraft}.
	\bigskip
	\caption{Recommended values of $\vec k$, for various scenario and parameter behavior for a given camera motion type.}
	\label{tab:presets}
\end{table}

\begin{figure*}
	\begin{subfigure}{\columnwidth}
		\centering
		\includegraphics{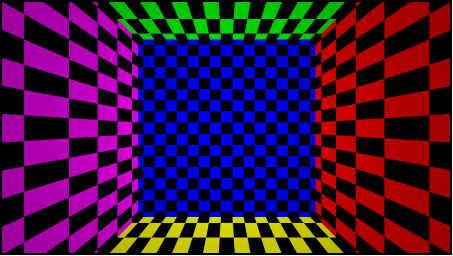}
		\caption{$k=1$, Rectilinear (standard).}
	\end{subfigure}
	\\[1em]
	\begin{subfigure}{\columnwidth}
		\centering
		\includegraphics{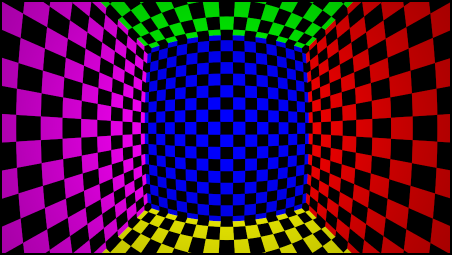}
		\caption{$\vec k=\big[\nicefrac{1}{2} \quad \nicefrac{22}{25}\big]$, first-person (anamorphic style).}
	\end{subfigure}
	\hfill
	\begin{subfigure}{\columnwidth}
		\centering
		\includegraphics{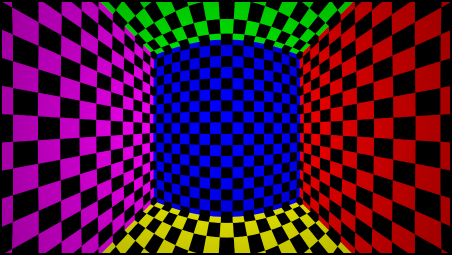}
		\caption{$\vec k=\big[\nicefrac{1}{2} \quad 1\big]$, panini.}
	\end{subfigure}
	\\[1em]
	\begin{subfigure}{\columnwidth}
		\centering
		\includegraphics{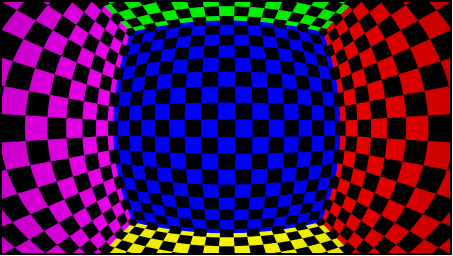}
		\caption{$\vec k=\big[\nicefrac{1}{2} \quad 0 \quad -\nicefrac{1}{2}\big]$, flying (asymmetrical).}
	\end{subfigure}
	\hfill
	\begin{subfigure}{\columnwidth}
		\centering
		\includegraphics{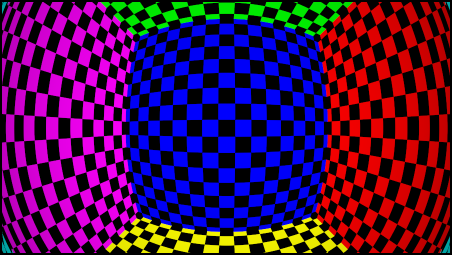}
		\caption{$\vec k=\big[-\nicefrac{1}{2} \quad \nicefrac{1}{2} \quad 0\big]$, racing (asymmetrical).}
	\end{subfigure}
	\caption{Example of various wide--angle ($\Omega_v\approx110\degree$) aximorphic-azimuthal projections with vignetting in $\nicefrac{4}{3}^2$ aspect-ratio. The checkerboard depicts a cube centered at the observation point, with each face colored according to the axis direction. Here, primary colors represent the positive axis, and neighboring complementary colors its negative equivalent (same as in the color-wheel), $\{Mg,Yl,Cy\}\mapsfrom-\ \{X,Y,Z\}\ +\mapsto\{R,G,B\}$.}
\end{figure*}

\section[Ray-map to ST-map]{Converting Ray-map to ST-map}

The ray/lens-map can be easily converted to the \emph{ST}-map format for distorting a rectilinear source image, provided that the maximum view angle $\Omega$ does not exceed or equal 180\degree.

\begin{subequations}
\begin{align}
	\begin{bmatrix}
		\vec a_x & \vec a_y
	\end{bmatrix}
	&=
	\begin{cases}
		\begin{bmatrix}
			1 & \frac{w}{h}
		\end{bmatrix}, &\text{if }\Omega_h
		\smallskip\\
		\begin{bmatrix}
			\frac{h}{w} & 1
		\end{bmatrix}, &\text{if }\Omega_v
	\end{cases}
\\
	\begin{bmatrix}
		\vec f_s \\
		\vec f_t
	\end{bmatrix}
	&=
	\frac{\cot\frac{\Omega}{2}}{2\hat G_z}
	\begin{bmatrix}
		\hat G_x \\
		\hat G_y
	\end{bmatrix}
	\begin{bmatrix}
		\vec a_x \\
		\vec a_y
	\end{bmatrix}
	+\frac{1}{2}, \qed
\end{align}
\end{subequations}
where $\vec a\in\real^2$ is the square-mapping vector for both the horizontal and vertical angle of view. Values $w$ and $h$ represent picture width and height, respectively. $\Omega<\pi$ is the angle of view. $\vec f\in[0,1]^2$ represent the final \emph{ST}-map vector. $\hat G\in[0,1]^3$ is the input viewing-ray map vector.

\section{Aximorphic Lens Distortion}

The presented perspective model can be used to mimic the real-world anamorphic lens and its effects. Effects such as \emph{disproportionate lens breathing} using focal length-based parametrization, which are unique to anamorphic photography \cite{Sasaki2017_5PilarsOfAnamorphic,Neil2004CineLens}.
Some additional lens-corrections may be added to the initial ray-map, to simulate more complex lens distortions and lens imperfections.

Below, an algorithm for aximorphic distortion of view coordinates is presented, which can be used as an input for viewing-ray map algorithm (equation \vref{eq:double spherical theta}). The algorithm is based on the \noun{Brown-Conrady} lens-distortion model \cite{Wang2008BrownConradyNewModelLensDistortion} in a division variant \cite{Fitzgibbon2001LensDivision}. It is executed on a view-coordinate $\vec v$, forming alternative vector $\vec v'$.

\begin{subequations}
\begin{align}
	\begin{bmatrix}
		\vec f_x \\
		\vec f_y
	\end{bmatrix}
	&=
	\underbrace{
		\begin{bmatrix}
			\vec v_x-\vec c_1 \\
			\vec v_y-\vec c_2
		\end{bmatrix}
	}_\text{cardinal offset \emph{a}}
	\\
	\begin{bmatrix}
		\vec\varphi_x \\
		\vec\varphi_y
	\end{bmatrix}
	&=
	\begin{bmatrix}
		\frac{\vec f^2_x}{\vec f^2_x+\vec f^2_y} &
		\frac{\vec f^2_y}{\vec f^2_x+\vec f^2_y}
	\end{bmatrix}\tran
	\\
	r^2 &= \vec f^2_x+\vec f^2_y
	\\
	\begin{split}
		\begin{bmatrix}
			\vec v'_x \\
			\vec v'_y
		\end{bmatrix}
	&=
		\begin{bmatrix}
			\vec f_x \\
			\vec f_y
		\end{bmatrix}
		\underbrace{
			\Bigg(
				\begin{bmatrix}
					1+\vec k_{x1}r^2+\vec k_{x2}r^4\cdots \\
					1+\vec k_{y1}r^2+\vec k_{y2}r^4\cdots
				\end{bmatrix}
				\cdot
				\begin{bmatrix}
					\vec\varphi_x \\
					\vec\varphi_y
				\end{bmatrix}
			\Bigg)^{-1}
		}_\text{radial aximorphic} \qed
	\\&\quad
		+
		\underbrace{
			\begin{bmatrix}
				\vec f_x \\
				\vec f_y
			\end{bmatrix}
			\left(
				\begin{bmatrix}
					\vec f_x \\
					\vec f_y
				\end{bmatrix}
				\cdot
				\begin{bmatrix}
					\vec p_1 \\
					\vec p_2
				\end{bmatrix}
			\right)
		}_\text{decentering}
		+
		\underbrace{
			r^2
			\begin{bmatrix}
				\vec q_1 \\
				\vec q_2
			\end{bmatrix}
		}_\text{thin prism}
		+
		\underbrace{
			\begin{bmatrix}
				\vec c_1 \\
				\vec c_2
			\end{bmatrix}
		}_\text{cardinal \emph{b}} \square
	\end{split}
	\label{eq:lens distortion}
	\\
	\begin{bmatrix}
		\vec v'_x \\
		\vec v'_y
	\end{bmatrix}
	&\mapsto
	\begin{bmatrix}
		\hat G_x \\
		\hat G_y \\
		\hat G_z
	\end{bmatrix}, \qed
\end{align}
\end{subequations}
where $\vec c_1,\vec c_2$ are the cardinal-offset parameters, $\vec q_1,\vec q_2$ are the thin-prism distortion parameters and $\vec p_1,\vec p_2$ are the decentering parameters. A set of $\vec k$ parameters define radial distortion for each aximorphic power axis. $\vec v$ is the input view-coordinate, and $\vec v'$ is the view coordinate with applied lens-transformation. $\vec\varphi\in[0,1]^2$ is the aximorphic interpolation weight, defined in section \vref{sec:primary-ray map}.


\section{Aximorphic Chromatic Aberration}

\begin{figure}
	\includegraphics{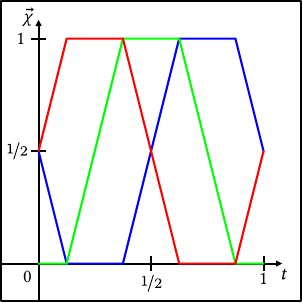}
	\caption{Mapping of $t\in[0,1]$ to spectral color $\vec\chi\in[0,1]^3$, for emulation of chromatic aberration. This is an output of periodic function, found in equation \eqref{eq:spectrum}. Distribution of the values ensures proper color order and sum-of-samples with guarantied neutral-white tint.}
	\label{fig:spectrum}
\end{figure}

A chromatic aberration effect can be achieved with multi-sample blur, where each sample layer is colored with the corresponding spectral-value \cite{Gilcher2015ChromaticAberration}. Presented periodic function for spectral color $\vec\chi$ produces samples that always add up to 1 (neutral white) when number of samples is even. It also exhibits the correct order of spectrum colors.
\begin{equation}
\label{eq:spectrum}
	\begin{bmatrix}
		\vec\chi_r \\
		\vec\chi_g \\
		\vec\chi_b
	\end{bmatrix}
	=
	\clamp_0^1
	\bigg(
		\frac{3}{2}-
		\big|
			4 \bmod
			\Bigg(
				t+\begin{bmatrix}
					\nicefrac{1}{4} \\
					0 \\
					\nicefrac{3}{4}
				\end{bmatrix}, 1
			\Bigg)-2
		\big|
	\bigg),
\end{equation}
where $\vec\chi\in[0,1]^3$ is the spectral-color value for position $t\in[0,1]$ (fee figure \vref{fig:spectrum} for more information).

Performing a spectral blur on an image involves the sum of multiple spectrum-colored layers. Scalar $t$ (here replaced by sample progress) should never reach 1, which ensures preservation of picture's white-balance. The number of samples $n$ must be an even number, and no less than 2 for a correct, neutral-white result.
\begin{equation}
	\begin{bmatrix}
		\vec f'_r \smallskip\\
		\vec f'_g \smallskip\\
		\vec f'_b
	\end{bmatrix}
	=
	\frac{2}{n}
	\sum_{i=0}^{n-1}
		\begin{bmatrix}
			\vec f_r \smallskip\\
			\vec f_g \smallskip\\
			\vec f_b
		\end{bmatrix}
		\underbrace{
			\clamp_0^1
			\bigg(
				\frac{3}{2}-
				\big|
					4 \bmod
					\Bigg(
						\frac{i}{n}+
						\begin{bmatrix}
							\nicefrac{1}{4} \\
							0 \\
							\nicefrac{3}{4}
						\end{bmatrix}, 1
					\Bigg)-2
				\big|
			\bigg)
		}_{\vec\chi\text{ periodic function}},
\end{equation}
where $n\in2\positive_1$ is the even number of samples for the chromatic aberration color-split.
$\vec f\in[0,1]^3$ is the current-sample position color-value. $\vec f'\in[0,1]^3$ is the final spectral-blurred color value.

The equation for spectral color $\vec\chi$ can be rewritten to a more optimized form, for parallel computation.
\begin{equation}
	\begin{bmatrix}
		\vec\chi_r \\
		\vec\chi_g \\
		\vec\chi_b
	\end{bmatrix}
	=
	\begin{bmatrix}
		\hfill \clamp_0^1\big(\nicefrac{3}{2}-|4t-1|\big) \smallskip\\
		\hfill \clamp_0^1\big(\nicefrac{3}{2}-|4t-2|\big) \smallskip\\
			  -\clamp_0^1\big(\nicefrac{3}{2}-|4t-1|\big) \\
	\end{bmatrix}
	+
	\begin{bmatrix}
		\hfill \clamp_0^1\big(4t-\nicefrac{7}{2}\big) \smallskip\\
		0 \smallskip\\
			 1-\clamp_0^1\big(4t-\nicefrac{7}{2}\big)
	\end{bmatrix},
\end{equation}
where $\vec\chi\in[0,1]^3$ is the spectral color at position $t\in[0,1]$. See the figure \vref{fig:spectrum} for visualization.

\subsection{Chromatic Aberration through Lens Distortion}

\begin{figure}
	\includegraphics{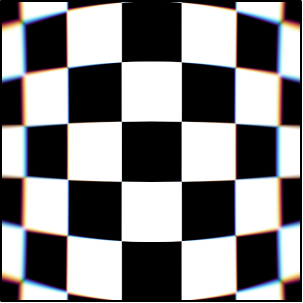}
	\caption{Example of aximorphic lens distortion with chromatic aberration, where $\vec k_{x1}=-0.25$, $\vec k_{y1}=0.04$, $d=0.5$, with 64--spectral samples.}
	\label{fig:lens distortion chromatic}
\end{figure}

Chromatic aberration can be integrated into lens distortion with spectral blurring, through the lens-transformation vector $\Delta\vec v$. Below, the equation for the spectral-blur displacement vector $\vec s$ is presented, calculated per sample at position $t$.
\begin{subequations}
\begin{align}
	\begin{bmatrix}
		\Delta\vec v_x \\
		\Delta\vec v_y
	\end{bmatrix}
	&=
	\begin{bmatrix}
		\vec v'_x-\vec v_x \\
		\vec v'_y-\vec v_y
	\end{bmatrix}
	\\
	\begin{bmatrix}
		\vec s_x \\
		\vec s_y
	\end{bmatrix}
	&= \Big( 1+\bigg(t-\frac{1}{2}\bigg)d \Big)
	\begin{bmatrix}
		\Delta\vec v_x \\
		\Delta\vec v_y
	\end{bmatrix},
\end{align}
\end{subequations}
where $\vec s\in\real^2$ is the spectral blur sample-offset vector at position $t\in[0,1)$. Value $d\in\real$ denotes the lens dispersion-scale.
For a visually pleasing result, additional blur pass can be applied, which direction is perpendicular to the $\vec s$ vector, and smaller in magnitude, as below.
\begin{equation}
	\begin{bmatrix}
		\vec s'_x \\
		\vec s'_y
	\end{bmatrix}
	=
	\frac{1}{4}
	\begin{bmatrix}
		-\vec s_y \\
		 \vec s_x
	\end{bmatrix},
\end{equation}
where $\vec s'$ is the second-pass blur direction vector.

Figure \vref{fig:lens distortion chromatic} presents the final effect of two-pass blur, where the first pass is spectral, along lens-distortion $\Delta\vec v$, and the second-pass (of quarter-magnitude) is perpendicular $\bot\Delta\vec v$. A combination of both adds a defocusing effect to the image distortion.

\section{Final Thoughts}

In this paper, a mathematical model for generating various asymmetrical aximorphic perspective projections has been provided, along with perception-driven distortion-design recommendations. In the model, each principal axis of the image resembles some azimuthal projection exactly, while regions in-between are transitional, creating a hybrid projection. This way, advantage can be taken of many projection types, to create an optimal, tailored view for a given specific scenario.
Such perception-driven parametrization enables the picture's geometry to adapt to the context, allowing dynamic adjustments on-the-fly in an artistically convincing manner.

In addition to aximorphic perspective, this paper presents vignetting effects and lens distortion with integrated chromatic aberration. The selection of these key features, enables almost complete digital lens simulation, for an immersive-imagery production.

Additional shader implementation of this technique can be found in the open-source \verb|PerfectPerspective.fx| shader available on \href{https://github.com/Fubaxiusz/fubax-shaders}{GitHub} or via \href{https://reshade.me/}{ReShade} platform.

\subsection[Prospects]{Prospects for Future Improvement}

In the future, wide choice of cinematic lenses may be totally replaced by a post-production technique or an in-camera special effect, performed on-stage.
Such technical solution would require several new technologies, like perhaps universal camera system, consisting of:
\begin{itemize}
	\item Universal lens:
	\begin{itemize}
		\item With parallax aberration correction and carefully mapped distortion.
	\end{itemize}
	\item Universal camera sensor:
	\begin{itemize}
		\item In correlation with the lens would produce distortion-free picture. Perhaps a light-field capturing sensor.
	\end{itemize}
	\item Universal perspective algorithm:
	\begin{itemize}
		\item Satisfied with this document.
	\end{itemize}
\end{itemize}

\begin{figure*}
	\begin{subfigure}[t]{3in}
		\includegraphics{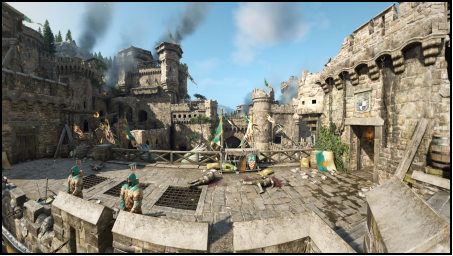}
		\footnotesize\emph{Panorama source:} Captured from \noun{For Honor} through Nvidia Ansel.
		\caption{$\vec k=[\nicefrac{1}{2} \quad 1]$, $f=0.6$, $\Omega_h\approx159\degree$, panini preset, which along vertical axis, preserves straight lines, and along horizon preserves proportions.}
		\label{fig:panini panorama}
	\end{subfigure}
	\hfil
	\begin{subfigure}[t]{3in}
		\includegraphics{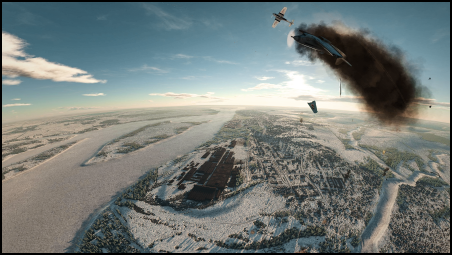}
		\footnotesize\emph{Panorama source:} Captured from \noun{War Thunder} through Nvidia Ansel.
		\caption{$\vec k=[\nicefrac{1}{2} \quad 0 \quad -\nicefrac{1}{2}]$, $f=0.6$, $\Omega_h\approx159\degree$, flying (asymmetrical), where bottom-half preserves distance, horizon preserves shape, and top-half preserves speed.}
		\label{fig:flying panorama}
	\end{subfigure}
	\\[1em]
	\begin{subfigure}[t]{3in}
		\includegraphics{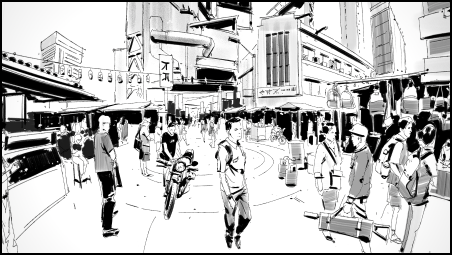}
		\footnotesize\emph{Panorama source:} ``China town'' by \noun{Jama Jurabaev}.
		\caption{$\vec k=[0 \quad 1 \quad \nicefrac{4}{5}]$, $\Omega_h=180\degree$, artistic projection where background architecture points straight-up. Attention is focused at the center figure by compression of periphery, and proper composition is achieved by expansion of the bottom field.}
		\label{fig:cartoon panorama}
	\end{subfigure}
	\hfil
	\begin{subfigure}[t]{3in}
		\includegraphics{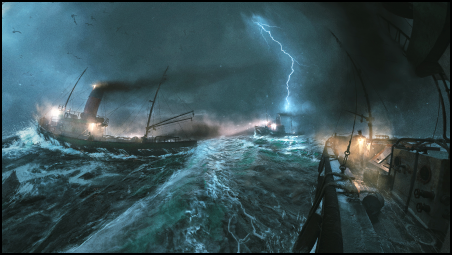}
		\footnotesize\emph{Panorama source:} ``The Lighthouse'' by \noun{Vladimir Somov}.
		\caption{$\vec k=[\nicefrac{1}{2} \quad 0 \quad \nicefrac{5}{8}]$, $\Omega_h=195\degree$, artistic projection with wide vertical field and limited bottom field, preserving proportion on the horizontal axis.}
		\label{fig:artistic panorama}
	\end{subfigure}
	\caption{Examples of super wide-angle views in aximorphic projection with natural vignetting. Mapped from various 360\textdegree\ panoramas in equirectangular projection.}
	\label{fig:panorama examples}
\end{figure*}


\begin{acks}

I would like to thank Anna and Stanley Sobczynski for their help, putting finishing touches on this document.


\end{acks}


\label{sec:reference}
\bibliographystyle{ACM-Reference-Format}
\bibliography{bibliography} 




\onecolumn
\pdfbookmark{Code Listings}{bm:listings}
\label{sec:code}

\begin{lstlisting}[
	% Bypass licence notification (repeated)
	firstline=8,
	caption={Mapping function for \emph{field of view} from \emph{focal length} and $k$ parameter in GLSL.},
	label={lst:fov}
]
float get_fov(float k, float focal)
{
	if      (k > 0.0) return degrees(2 * atan(k / focal) / k); // stereographic, rectilinear projections
	else if (k < 0.0) return degrees(2 * asin(k / focal) / k); // equisolid, orthographic projections
	else   /*k == 0*/ return degrees(2 / focal);               // equidistant projection
}
\end{lstlisting}

\begin{lstlisting}[
	% Bypass licence notification (repeated)
	firstline=8,
	caption={Mapping function for \emph{focal length} from \emph{field of view} and $k$ parameter in GLSL.},
	label={lst:focal}
]
float get_rcp_focal(float halfOmega, float radiusOfOmega, float k)
{
	if      (k > 0.0) return tan(k * halfOmega) / (radiusOfOmega * k); // stereographic, rectilinear projections
	else if (k < 0.0) return sin(k * halfOmega) / (radiusOfOmega * k); // equisolid, orthographic projections
	else   /*k == 0*/ return         halfOmega  /  radiusOfOmega;      // equidistant projection
}
\end{lstlisting}

\begin{lstlisting}[
	% Bypass licence notification (repeated)
	firstline=8,
	caption={Mapping function for $\theta$ angle of azimuthal projection in GLSL.},
	label={lst:theta}
]
float get_theta(float radius, float rcpFocal, float k)
{
	if      (k > 0.0) return atan(radius * rcpFocal * k) / k; // stereographic, rectilinear projections
	else if (k < 0.0) return asin(radius * rcpFocal * k) / k; // equisolid, orthographic projections
	else   /*k == 0*/ return      radius * rcpFocal;          // equidistant projection
}
\end{lstlisting}

\begin{lstlisting}[
	% Bypass licence notification (repeated)
	firstline=8,
	caption={Mapping function for azimuthal projection radius from $\theta$ angle in GLSL.},
	label={lst:radius}
]
float get_radius(float theta, float rcpFocal, float k)
{
	if      (k > 0.0) return tan(k * theta) / (rcpFocal * k); // stereographic, rectilinear projections
	else if (k < 0.0) return sin(k * theta) / (rcpFocal * k); // equisolid, orthographic projections
	else   /*k == 0*/ return         theta  /  rcpFocal;      // equidistant projection
}
\end{lstlisting}

\begin{lstlisting}[
	% Bypass licence notification (repeated)
	firstline=8,
	caption={Function for vignetting mask from projection radius and incidence angle $\theta$ in GLSL.},
	label={lst:vignette}
]
float get_vignette(float theta, float radius, float rcpFocal)
{ return sin(theta) / (radius * rcpFocal); }
\end{lstlisting}

\begin{lstlisting}[
	% Bypass licence notification (repeated)
	firstline=8,
	caption={Function for aximorphic interpolation $\phi$ weights in GLSL.},
	label={lst:phi}
]
vec2 get_phi_weights(vec2 viewCoord)
{
	viewCoord *= viewCoord; // squared vector coordinates
	return viewCoord / (viewCoord.x + viewCoord.y); // $[\cos^2\phi \quad \sin^2\phi]$ vector
}
\end{lstlisting}

\begin{lstlisting}[
	% Bypass licence notification (repeated)
	firstline=8,
	caption={Mapping function for chromatic aberration spectrum, from hue value in GLSL.},
	label={lst:lens}
]
vec3 spectrum(float hue)
{
	hue *= 4.0;
	vec3 hueColor;
	hueColor.rg = hue - vec2(1.0, 2.0);
	hueColor.rg = clamp(1.5 - abs(hueColor.rg), 0.0, 1.0);
	hueColor.r += clamp(hue - 3.5, 0.0, 1.0);
	hueColor.b  = 1.0 - hueColor.r;
	return hueColor;
}
\end{lstlisting}

\begin{lstlisting}[
	% Bypass licence notification (repeated)
	firstline=7,
	caption={Aximorphic and other lens distortion functions in GLSL.},
	label={lst:lens}
]
// Radial distortion-division model (add to coordinates)
vec2 radial(float radius2, vec2 viewCoord, vec2 kx, vec2 ky)
{
	float radius4 = radius2 * radius2; // $r^4$
	vec2 phiWeights  = viewCoord * viewCoord;
	     phiWeights /= phiWeights.x + phiWeights.y;
	vec2 kVec = vec2(
			1.0 + kx[0] * radius2 + kx[1] * radius4,
			1.0 + ky[0] * radius2 + ky[1] * radius4
		);
	return viewCoord / dot(kVec, phiWeights) - viewCoord;
}

// Decentering distortion model (add to coordinates)
vec2 decentering(vec2 viewCoord, vec2 p)
{ return viewCoord * dot(viewCoord, p); }

// Thin prism distortion model (add to coordinates)
vec2 thinPrism(float radius2, vec2 q)
{ return radius2 * q; }
\end{lstlisting}

\end{document}